\documentclass[aip,preprint,showpacs, secnumarabic, graphics, floatfix, nofootinbib, tightenlines, nobibnotes,aps,prl,10pt]{revtex4-1}
\usepackage{graphicx}
\usepackage[utf8]{inputenc}
\usepackage[T1]{fontenc}
\usepackage[english]{babel}
\usepackage{array,multirow,makecell}
\usepackage{amsmath}
\usepackage{amssymb}
\usepackage{mathrsfs}
\usepackage{amsmath}
\usepackage[squaren, Gray, cdot]{SIunits}
\usepackage{,amsfonts}
\usepackage{fancyhdr}
\usepackage{colortbl}
\usepackage[nottoc, notlof, notlot]{tocbibind}
\usepackage{vmargin}
\usepackage[colorlinks = true,linkcolor = blue, urlcolor  = blue,citecolor = blue,anchorcolor = blue]{hyperref}
\usepackage{natbib}
\usepackage{float}
\usepackage{url}

\usepackage[textsize=small]{todonotes}

\setmarginsrb{2cm}{0.5cm}{2cm}{3cm}{0cm}{1cm}{2cm}{1cm}
\makegapedcells
\newcolumntype{R}[1]{>{\raggedleft\arraybackslash }b{#1}}
\newcolumntype{L}[1]{>{\raggedright\arraybackslash }b{#1}}
\newcolumntype{C}[1]{>{\centering\arraybackslash }b{#1}}

\def\be{\begin{equation}}
\def\ee{\end{equation}}
\def\bea{\begin{eqnarray}}
\def\eea{\end{eqnarray}}

\newcommand{\ind}[1]{_{\mathrm{#1}}}

\begin{document}

\title{\textbf{\LARGE{Memory effects in the ion conductor Rb$_{2}$Ti$_{2}$O$_{5}$}}}
\author{Rémi Federicci}
\author{Stéphane Holé}
\affiliation{LPEM, ESPCI Paris, PSL Research University, CNRS, Sorbonne Université, 75005 Paris, France}

\author{Vincent Démery}
\affiliation{Gulliver, ESPCI Paris, PSL Research University, CNRS, 10 rue Vauquelin, 75005 Paris, France}
\affiliation{Univ Lyon, ENS de Lyon, Univ Claude Bernard Lyon 1, CNRS, Laboratoire de Physique, F-69342 Lyon, France}

\author{Brigitte Leridon$^1$}
\date{\today}

\begin{abstract}

Recent studies on Rb$_2$Ti$_2$O$_5$ crystals have demonstrated remarkable electrical properties. This material exhibits colossal electrical polarization between 200\,K and 330\,K. In the present work, we report on the observation of memory effects in Rb$_2$Ti$_2$O$_5$ due to charge accumulation and we discuss the genuine memristive character of this material. 
An analytical model is proposed for the system, which takes into account the ionic diffusion and ionic migration and is in good agreement with the observed volatile memristive properties of the material.

\end{abstract}

\maketitle
\section{Introduction}

Memristive systems are ubiquitous in nature, from the simple tungsten filament \cite{Prodromakis:2012kza} to the complex synaptic junction in neuronal systems \cite{MemriBible}.
Predicted by Chua in 1971\cite{Chua1}, and claimed to be discovered at the nanoscale in 2008 by Strukov and coworkers \cite{Strukov}, the memristor is the electronic component whose state equation is expressed by a linear dependence between the charge $q$ and the flux $\Phi=\int V(t) dt$. 
The consequence is that $V=R(q)I$ where $R$ (the memristance) depends on the total charge $q$ received by the device. 
Therefore the history of the currents and voltages experienced by the material impacts its current internal state, giving rise to memory effects \cite{Chua3, Chua2}.
In the recent years, the research about memristors and more generally about memristive, memcapacitive and meminductive systems---where respectively $C$ or $L$ are a function of charge or flux---has become a topic of huge interest especially in regards of  neuromorphic implementation \cite{NatureMaterial-Neuromorphiccomputing, NatureTechno-Neuromorphiccomputing, Nanoscale-Neuromorphiccomputing, MemriNeurone, 2017-Cario-LIFNeuron} and data storage  (ReRAM) \cite{Chua3,Waser:hw,Sawa:2008vx}. 

Recent studies have shown remarkable electrical transport properties in Rb$_{2}$Ti$_{2}$O$_{5}$ (RTO) crystals or ceramics \cite{Federicci:2017csa}.
An ionic conductivity up to $10^{-3}$\,S\,cm$^{-1}$ between 200\,K and 330\,K was found, making the RTO a superionic conductor in this temperature range. 
Ionic transport in this highly electronically-insulating material creates accumulation of ions at the edges of the system, giving rise to an unscreened macro-dipole with a polarization larger than $0.1$\,C\,cm$^{-2}$ and a colossal equivalent relative permittivity of about $10^8$ at low frequency \cite{Federicci:2017csa}.  
The possibility of a conventional ferroelectric phase associated to a structural phase transition was completely discarded by Raman spectroscopy and XRD measurements. Indeed, RTO structure was found to remain consistent with the centrosymmetric C2$/$m space group between 10\,K and 450\,K \cite{2-2016-RTO-Structure}. 

In this article, we demonstrate intrinsic memory effects in bulk RTO at room temperature.
We also discuss the intrinsic memristor character of the material and the volatile nature of the memory effects. 
In addition,  a simple analytical model based on the electrolytic nature of the material is presented. 
The I-V curves calculated from the model are in very good agreement with the experimental measurements, and reproduce the observed non-transverse memristive behavior, further confirming the ionic mobility origin of these effects.

\section{Sample preparation and experimental set-up}

The RTO crystal synthesis was performed by melting TiO$_2$ and MNO$_3$ powders in a Pt crucible \cite{Federicci:2017csa} and characterization was performed as described in previous work \cite{Federicci:2017csa}. 
The crystals, which size is typically  1\,mm $\times$ 300$\,\mu$m $\times$ 300\,$\mu$m, were then extracted from the synthesis batch and silver wires were glued with carbon paste at both ends of the elongated platelet-type crystals.  The entire system was then annealed at 400\,K under room atmosphere to polymerize the carbon paste. Once connected, in order to activate the electrical transport properties\cite{Federicci:2017csa}, the crystals were annealed under vacuum at 400\,K for two hours.

Due to the very high impedance of our samples, we used a two-contact geometry. 
Voltage biased I-V curves were acquired using a Lecroy LT224 oscilloscope for measuring the voltage drop across a resistance of 200\,k$\Omega$ in series with the sample, and a Stanford SI1200 Lock-In was used as voltage source. For triangular voltage measurements and current relaxations a Keithley 6517b electrometer was used to measure the current while applying a voltage. All the transport measurements presented in the present article were performed under room pressure and room temperature conditions.\newline

\section{Electrical measurements}

\subsection{Memristive behavior}

\begin{figure}
\includegraphics[width=8.5cm, trim = 1.1cm 1cm 1cm 1.5cm, clip]{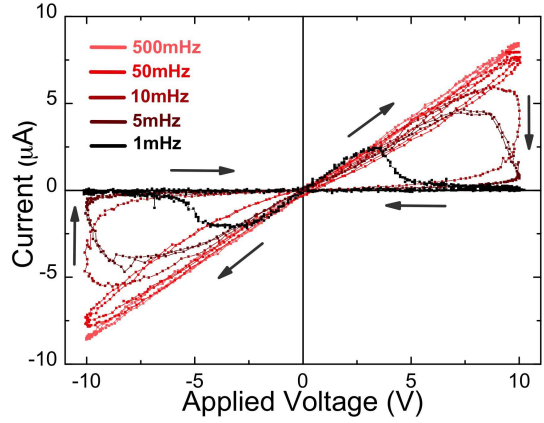}
\caption{I-V curves for  Rb$_2$Ti$_2$O$_5$ single crystal (sample A) under room conditions for frequencies ranging from 1\,mHz to 500\,mHz. The sample is a single crystal which dimensions are 1000\,$\mu$m $\times$ 300\,$\mu$m $\times$ 300\,$\mu$m. This crystal demonstrates genuine memristive behaviour, with an almost ideal ``memristor'' character according to the criteria defined by Adhikari et al.\cite{Adhikari1} (see text). }
\label{RTO-IV-Memoire}
\end{figure}

Voltage biased I-V curves were acquired for different frequencies ranging from 1\,mHz to 500\,mHz.
Typical results are displayed in Fig.~\ref{RTO-IV-Memoire}.
The results show non linear non transverse --- the two branches do not
cross at the origin --- hysteretical loops that are pinched in zero.
The area of the lobes increases with frequency from 1\,mHz to 10\,mHz, and decreases above 10\,mHz.
When the frequency of the applied voltage reaches 500\,mHz, the I-V curve shows an ohmic behavior with a single-valued conductivity of about $10^{-4}$\,S\,cm$^{-1}$.

For a system to be qualified as an ideal  memristor, three conditions must be met \cite{Adhikari1}. The first is that the I-V curve displays a pinched (either transverse or non-transverse) hysteretical loop, \textit{i.e.}, $I=0$ when $V=0$. 
The second condition is that the lobe area enclosed by the I-V curve decreases with increasing frequency.  And the third condition is that at infinite frequency, the I-V curve should be single-valued.
The first criterium is met since to a first approximation the loops are pinched in zero.  
We discuss later some experimental cases when the loops are not exactly pinched in zero and when the I-V lobes are not exactly restricted to the first and third quadrants.
The second criterion is not exactly fulfilled since the area of the butterfly lobes monotonously decreases with frequency only at high frequencies. 
The  third criterion for ideal memristor is fulfilled since an ohmic regime is observed at high frequency. 

These findings demonstrate that a properly annealed crystal of RTO connected to conducting wires is a genuine memristive device, close to an ideal memristor, at low frequencies and at room temperature. 

\subsection{Memory-volatile character}

As a matter of fact, as can be seen from the arrows drawn in Fig. \ref{RTO-IV-Memoire}, the I-V curves are of a non-transverse memristor, which is, for instance,  different from all known examples of resistance switching memory devices based on ion conductors\cite{Waser:hw}.
Consequently, it was observed in some of the samples, either single crystals or polycrystals, that the loop was slightly opened (not fully pinched)---which violates the first criterion. 
In this case when the bias voltage goes from positive to zero, a negative current is created at zero voltage, releasing charge into the outer circuit. 
We refer to this current as the ``relaxation current''.
The necessary conditions to produce this cycle opening and the relaxation current (which varies from sample to sample) are not fully understood, but seem to be related to the annealing conditions, the presence of defects in the sample and possibly the nature of the electrodes.
In this case, the device can be modeled by adding a memcapacitive component to the memristive one, and the system then partially behaves like a supercapacitor.

\begin{figure}
\includegraphics[width=8.5cm, trim = 1.1cm 0.8cm 1cm 1.5cm, clip]{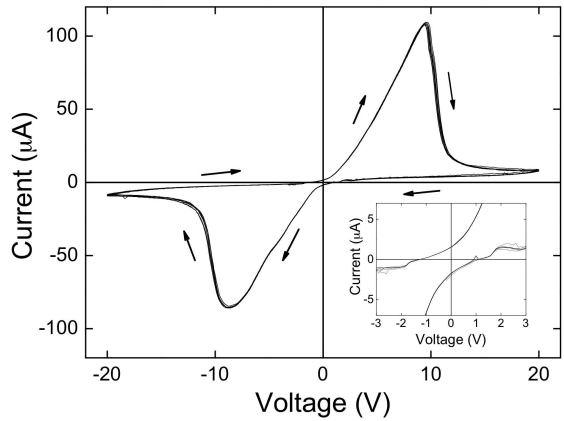}
\caption{I-V curves acquired at 1.2\,mHz on  Rb$_2$Ti$_2$O$_5$ (sample B) under room atmosphere. The sample is a polycrystal whose dimensions are 1500\,$\mu$m $\times$ 700\,$\mu$m $\times$ 800\,$\mu$m. There are eight I-V cycles that appear superimposed demonstrating the excellent reproducibility of the measurements.}
\label{RTO-IV-Memoire2}
\end{figure}

Fig.~\ref{RTO-IV-Memoire2} displays a set of eight I-V curves measured on the polycrystalline sample B where the relaxation currents can be observed. 
The 
frequency of the applied field is about 1.2\,mHz. The perfect coincidence between these eight curves demonstrates the reproducibility of the measurements. 
The inset in Fig.~\ref{RTO-IV-Memoire2} shows a close-up around the origin in order to outline the relaxation current.

\begin{figure*}[t]
\center
\includegraphics[width=16cm, trim = 4cm 2cm 4cm 0.5cm, clip]{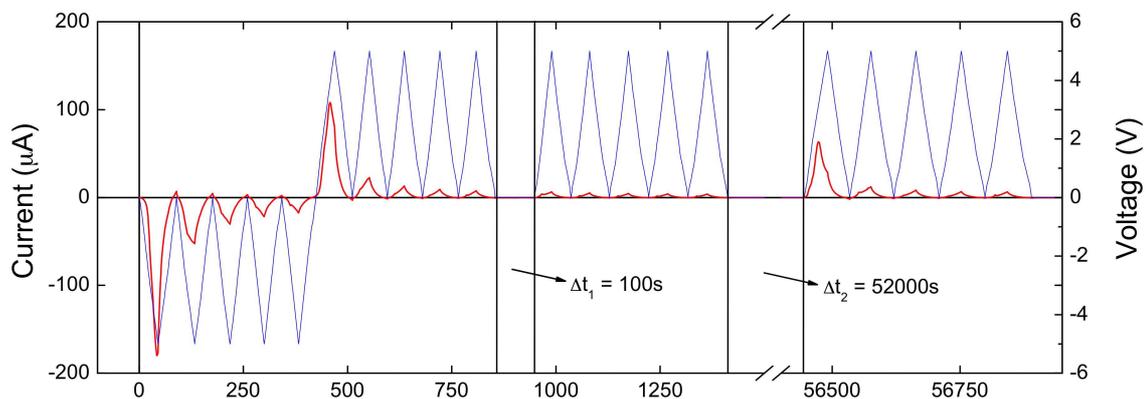}
\caption{Evolution of the current measured on a Rb$_2$Ti$_2$O$_5$ sample at room conditions as function of applied voltage. The blue curve is for the triangular applied voltage and the red curve is the measured current. The crystal dimensions are 3500\,$\mu$m $\times$400 \,$\mu$m $\times$ 1500\,$\mu$m.}
\label{RTO-profile-Memoire}
\end{figure*}

In order to better evidence the memristive properties of RTO, a triangular voltage signal was used to study the evolution of the internal state of the material as a function of time. 
The measured current (red) and the applied voltage (blue) are displayed as a function of time in Fig.~\ref{RTO-profile-Memoire}. A serie of 5 negative cycles between $-5$\,V and 0\,V was performed, followed by a series of 5 positive cycles between 0\,V and 5\,V. These alternatively negative and positive cycles show how the conducting properties of RTO are impacted by the voltage history.  

First, during the series of negative cycles, the peak current amplitude decreases regularly in absolute value between the first and the last cycle and the peak current for the last cycle is observed to be almost 10 times smaller than for the first one. Once the series of positive cycles starts, the conductivity of the material is suddenly increased, being almost five times higher than during the previous last negative cycle. 
During this series of positive cycles, the conductivity is also observed to decrease in a similar way as for the series of negative cycles. 

After the set of alternatively negative and positive cycles, the applied voltage was reduced to zero during a time $\Delta t$ for a relaxation to occur. 
After this elapsed time, a series of 5 positive cycles was performed again to measure the internal conducting state of the RTO. The first cycle of the series exhibits the highest conductivity which is used to estimate the internal conducting state. The process was performed with different elapsed times and the current intensity of the first positive cycle was observed to increase with increasing elapsed time. This measurement shows that the conductivity of the RTO decreases with the increasing flux and that a slow leaky process exists when the applied voltage is zero which tends to restore the conductivity, attesting to the volatile nature of the memory effects at room temperature in this material.

\subsection{Origin of the memory effects}

Memory effects in devices comporting ion conductors are largely documented.  Both unipolar and bipolar I-V characteristics are reported, depending on the microscopic nature of the underlying mechanims \cite{Waser:hw}. Memory effects are also largely studied in perovskite-based devices \cite{Sawa:2008vx}.

A very commonly reported  mechanism  is the creation of microscopic conduction channels through the application of a high voltage, which provokes a local breakdown of the insulator \cite{Pagnia:1988,Chudnovskii:1996}. After application of a threshold voltage, these channels are destroyed because of local Joule heating. This gives a very typical I-V curve which depends on the compliance current, with an either bipolar or unipolar switching type.  In any case the I-V curve is always transverse,\textit{ i.e.} the two branches cross around the origin. This is not what is observed in our system. 

A second possible mechanism relies on redox processes associated with cation electromigration. In this case, the process involves the oxidation of a metallic electrode such as Ag, the migration of the formed cations such as, \textit{e.g.}, silver ions  Ag$^+$ through the ion conductor and the deposition at  an inert electrode whose role is simply to provide electrons.  
This process, which creates silver dendrites, is reversible and leads to switching behavior, which was first reported in 1976 by Hirose \cite{Hirose:1976}.
In our case, the nature of the wires (Ag or Au) was varied without affecting the results. 
Gold contacts were evaporated on the cystals and used to perform the same measurements, leading to identical I-V curves so the role of Ag$^+$ ions can be excluded. 

A third reported mechanism is electrochemical migration of oxygen ions. Such mechanism was observed for example in STO thin films \cite{szot:2006}. It is very tempting indeed to consider such a mechanism in the present system, since we know that ionic motion is present, and the preparation process (annealing under oxygen-poor atmosphere and application of large voltage) reported for STO is very similar to the one applied on RTO. 
However, again, the I-V curves reported  in STO systems  \cite{Beck:2000ji, Watanabe:2001ju, szot:2006} are transverse in contradiction to the present observation. 
The direction of variation of the cycle in the first quadrant  is opposite to the one reported in Fig.~\ref{RTO-IV-Memoire}.  

As a matter of fact a notable difference between the two systems is present: in the above-reported mechanism, the mobile oxygen ions experience charge transfer through redox reaction at the anode interface. This means that the following reaction takes place,

\begin{equation}
\mathrm{O}^{2-}\to 2\mathrm{e}^- + \frac{1}{2} \mathrm{O}_2,
\end{equation}
and a production of oxygen gas is observed at the anode. In this case, ions are not accumulated and there is no building up of charge at the interface.  This is in contradiction to what is observed in the present system where ions remain accumulated,  a giant polarization (of the order of $0.1$\,C\,cm$^{-2}$) is created \cite{Federicci:2017csa}, and where no particular gas production is observed at the anode. 
Moreover, the phenomenon is also observed under helium atmosphere, which discards a possible role played by oxygen gas production and absorption.

In order to explain the observed I-V curve, we therefore developped a simple model taking into account the ion mobility, ion accumulation, and supposing a total absence of charge transfer at the interface.  This means that we assumed that redox reaction are completely absent at the interface where the ions accumulate and that for some reason remaining to be elucidated if the mobile ions are oxygen ions O$^{2-}$, which is very likely, they cannot escape the material and transform into oxygen gas.

\section{Ionic electromigration model}

\subsection{Model}\label{}

In order to better understand the microscopic mechanisms at play in this material, we compare the experimental results with a simple model of ionic migration.
First, we give the general relations satisfied by this kind of system, and then make additional hypothesis to arrive at a system of equations that can be integrated numerically.

We assume that all the quantities only depend on the coordinate $x\in[0,L]$, where $L$ is the length of the sample, and that all the fields are oriented along this direction.
The Poisson's law thus reads
\begin{equation}
\partial_x E(x,t)=\frac{\rho(x,t)}{\epsilon},
\end{equation}
where $E$ is the electric field and $\rho$ the charge density. 
The evolution of the charge density is given by the current $j$:
\begin{equation}
\partial_t\rho (x,t)=-\partial_x j(x,t).
\end{equation}

The electric field at the boundary $x=L$ is related to the surface charge density $\sigma(t)$ of the right electrode through
\begin{equation}
E(L,t) = -\frac{\sigma(t)}{\epsilon}.
\end{equation}
The potential difference between the electrodes can be written as
\begin{equation}
V(t)=- \int_0^L E(x,t)dx = \frac{L\sigma(t)}{\epsilon} + \frac{1}{\epsilon}\int_0^L x\rho(x,t)dx,
\end{equation}
where we have integrated by parts and used Poisson's law.
This equation shows how imposing $V(t)$ sets the charge on the electrodes, given the distribution of charges.
Moreover assuming that there are no charges entering or leaving the sample at the electrodes, 
\begin{equation}
j(0,t) = j(L,t) = 0,
\end{equation}
we can derive the previous relation and get for the current
\begin{equation}\label{eq:decomp_current}
I(t)=A \frac{d\sigma}{dt}(t) = \frac{A\epsilon}{L}\frac{dV}{dt}(t) - \frac{A}{L}\int_0^L j(x,t)dx,
\end{equation}
where $A$ is the area of the electrodes, or equivalently the area of the section of the sample.
The current thus consists in two terms (Eq.~(\ref{eq:decomp_current})): the capacitive current $I\ind{cap}(t)$ and the ionic current $I\ind{ion}(t)$.

We now make further assumptions to describe ionic migration in the sample. 
We assume that there is a single species of mobile ions, carrying a charge $q$, and we denotes its number density $n(x,t)$.
We further assume that the material is neutral when the mobile ions are uniformly distributed in the material, meaning that there is a uniform background of non-mobile charges, with number density $\bar n$.
The charge density thus reads:
\begin{equation}
\rho(x,t)=q[n(x,t)-\bar n].
\end{equation}
We make the hypothesis that the mobile ions move according to the Nernst-Planck equation:
\begin{align}
\partial_t n(x,t) & = -\partial_x j_n(x,t),\\
j_n(x,t) & = -D\partial_x n(x,t) + \mu E(x,t) n(x,t),\label{eq:nernst_planck}
\end{align}
where $\mu$ is the mobility of the mobile ions, and $D$ is their thermal diffusion coefficient; they are related through the Einstein relation $D=k_B T q^{-1}\mu$.
The ionic flux is related to the current through $j(x,t)=qj_n(x,t)$.

Finally, in order to introduce the caracteristic amplitude $V_0$ and frequency $\omega$ of the driving, we write it as
\begin{equation}
V(t)=V_0f(\omega t),
\end{equation}
where $f(u)$ is a $2\pi$-periodic function with absolute maximum 1.

\subsection{Dimensionless parameters and equations}\label{}

The model contains many parameters; in order to understand its behavior, we introduce the following dimensionless quantities
\begin{align}
\tilde x & = \frac{x}{L},\\
\tilde t & = \frac{\mu q \bar n}{\epsilon}t,\\
\tilde n & = \frac{n}{\bar n},\\
\tilde E & = \frac{\epsilon}{q\bar n L}E,\\
\tilde \sigma & = \frac{\sigma}{q\bar n L},\\
\tilde I & = \frac{\epsilon I}{\mu q^2\bar n^2LA}.
\end{align}
These quantities allow to rewrite the equations as
\begin{align}
\partial_{\tilde t}\tilde n (\tilde x, \tilde t) & = -\partial_{\tilde x} \tilde j_n(\tilde x, \tilde t),\label{eq:adim_first}\\
\tilde j_n(\tilde x, \tilde t) & = \tilde E(\tilde x, \tilde t)\tilde n(\tilde x, \tilde t)- \tilde T\partial_{\tilde x}\tilde n(\tilde x, \tilde t),\\
\partial_{\tilde x}\tilde E(\tilde x, \tilde t) & = \tilde n(\tilde x, \tilde t)-1,\\
\tilde E(0,\tilde t) & = \tilde E(1,\tilde t) = -\tilde\sigma(\tilde t),\\
\tilde V(t) & = \tilde \sigma(\tilde t)+\int_0^1\tilde x \left[\tilde n (\tilde x, \tilde t)-1 \right]d\tilde x,\\
\tilde I(\tilde t) & = \frac{d\tilde V}{d\tilde t}(\tilde t)-\int_0^1\tilde j_n(\tilde x, \tilde t)d\tilde x,\\
\tilde V(\tilde t) & = \tilde V_0 f(\tilde \omega\tilde t).\label{eq:adim_last}
\end{align}
These equations depend only on three dimensionless parameters, which determine the behavior of the system:
\begin{align}
\tilde V_0 & = \frac{\epsilon V_0}{q\bar n L^2},\\
\tilde T & = \frac{\epsilon k_BT}{q^2\bar n L^2},\\
\tilde \omega & = \frac{\epsilon\omega}{\mu q\bar n}.
\end{align}

Since the parameters $\bar n$, $\mu$ and $\epsilon$ are not known, the dimensionless parameters cannot be computed directly.
However, the ratio $\tilde T/\tilde V_0$ can be estimated for the experiment corresponding to Fig.~\ref{RTO-IV-Memoire} at room temperature (300\,K), assuming that the mobile ions are oxygen ions carrying two elementary charges, and a 10\,V voltage drop:
\begin{equation}
\frac{\tilde T}{\tilde V_0} = \frac{k_B T}{qV_0} \simeq 10^{-3}.
\end{equation}

Furthermore, the ohmic behavior observed at high frequency in Fig.~\ref{RTO-IV-Memoire} can be exploited. 
The order of magnitude of the capacitive current in Eq.~(\ref{eq:decomp_current}) is
\begin{equation}
I\ind{cap}\sim \frac{A\epsilon\omega V_0}{L},
\end{equation}
and the ionic current may be estimated by assuming that the charge displacement is small, so that the effect on the electric field is weak, hence
\begin{equation}
I\ind{ion}\sim \frac{A\mu q\bar n V_0}{L}.
\end{equation}
In the experiment corresponding to Fig.~\ref{RTO-IV-Memoire}, a triangular signal is imposed, so that the capacitive current, which is proportional to the derivative of the imposed voltage, should be constant.
This constant is not observed, meaning that the capacitive current is much smaller than the ionic current, $I\ind{cap}\ll I\ind{ion}$.
Hence, the measured current is the ionic current, which allows to estimate the ionic conductivity
\begin{equation}
\mu q \bar n = \frac{L I\ind{ion}}{A V_0}\simeq 10^{-4}\,\mathrm{S}\,\mathrm{cm}^{-1}.
\end{equation}
We can check the consistency of our relation by evaluating the ratio of the currents at the highest frequency, which corresponds to the dimensionless frequency:
\begin{equation}\label{eq:est_omegat}
\tilde\omega 
= \frac{I\ind{cap}}{I\ind{ion}}
= \frac{\epsilon \omega}{q \mu\bar n}\simeq 5\,10^{-10}\epsilon_r,
\end{equation}
where $\epsilon_r$ is the relative permittivity of the medium.
Note that this permittivity is evaluated without the effect of the mobile charges, and should thus be of order one, so that the parameter $\tilde \omega$ should be very small.

\subsection{Numerical integration}\label{}

\begin{figure}
\begin{center}
\includegraphics[]{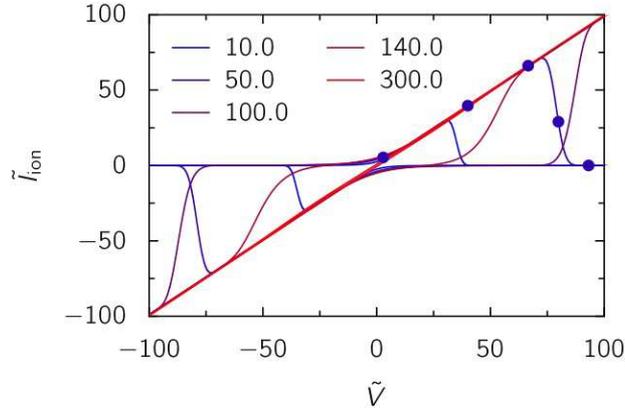}
\end{center}
\caption{Ionic current versus voltage drop in the Nernst-Planck-Poisson model Eqs.~(\ref{eq:adim_first}--\ref{eq:adim_last}) with a triangular driving for $\tilde V_0=100$, $\tilde T=0.1$ and different values of $\tilde \omega$ given in the legend.
The circles indicate the positions where the charge distribution is plotted in Fig.~\ref{fig:charge_dist} for $\tilde \omega=50$.}
\label{fig:simulations}
\end{figure}

We now turn to the numerical integration of the dimensionless equations (\ref{eq:adim_first}-\ref{eq:adim_last}).
We define the value of the density on a lattice and use a finite differences scheme that conserves the number of ions.
The ionic current $\tilde I\ind{ion}$ is plotted as a function of the voltage drop $\tilde V$ for a triangular signal with $\tilde V_0=100$, $\tilde T=0.1$ and different frequencies in Fig.~\ref{fig:simulations}.
We note the close resemblance with Figs.~\ref{RTO-IV-Memoire} and \ref{RTO-IV-Memoire2}.
The simulations reproduce several features of the experimental findings, such as the non monotonous variation of the cycle area with frequency and the existence of relaxation currents.

\begin{figure}
\begin{center}
\includegraphics[]{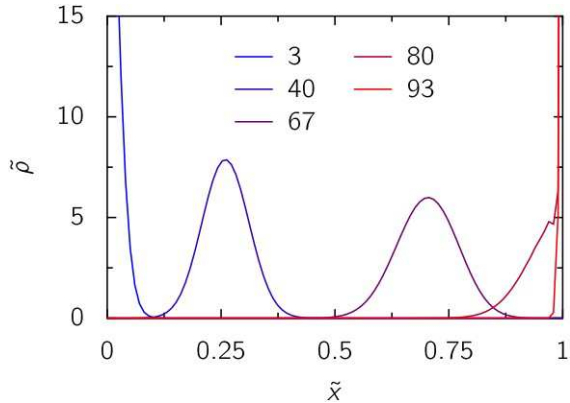}
\end{center}
\caption{Mobile charge distribution in the sample for $\tilde V_0=100$, $\tilde T=0.1$ and $\tilde\omega=50$ along the cycle. The values of $\tilde V$ are indicated in the legend, and correspond to the circles in Fig.~\ref{fig:simulations}.}
\label{fig:charge_dist}
\end{figure}

In order to get insight onto the physical mechanism responsible for the current drop during the cycles, we plot the charge distribution at different points of the cycle for $\tilde\omega=50$ in Fig.~\ref{fig:charge_dist}.
We observe that during a cycle, the mobile charges move as a whole from the left electrode to the right one, and spread slightly before hitting the right electrode.
The current drop indeed occurs when the charges meet the right boundary of the system.

\subsection{Discussion of the model}\label{}

Because of the diffusion force coming from the ionic concentration gradient, the ions motion is getting slower and slower.
This regime is nothing but the appearance of the Warburg diffusion \cite{Warburg1, Warburg2}, caused by the ionic accumulation and already observed in the low-frequency behavior of  the dielectric constant  \cite{Federicci:2017csa}.
Before the mobile ions begin to feel the edges of the sample they are in a free motion regime with a conductivity of $10^{-4}$\,S\,cm$^{-1}$ related to the ionic conductivity \cite{Federicci:2017csa}. 

However, important discrepancies remain between the model and the experimental results.
First, the value of the dimensionless parameter $\tilde\omega$ used in the model, $10\leq\tilde\omega \leq 300$, is incompatible with the experimental estimation for a reasonable value of the relative permittivity of the medium, Eq.~(\ref{eq:est_omegat}).
Second, we plot only the ionic current in the model, and discard the capacitive current.
Yet, for the values of the dimensionless parameters used, the capacitive current is much larger than the ionic current.
A possible explanation could be a bad choice of dimensionless parameters, but we could not recover the experimental behavior for other values of the parameters.
A key parameter is $\tilde V_0$, which compares the voltage drop to the mobile charges available in the system. 
The current drop at low frequency is observed only when $\tilde V_0\gg 1$, when the voltage drop is able to drive all the mobile ions to a boundary of the system.
We conclude that this model contains the basic ingredients to explain the experimental behavior, but misses some important physical mechanism.
Identifying such mechanism remains an open question and will certainly provide further insight onto the nature of ionic migration in such system.

The fact that the system is found to not strictly obey the first criterion for memristors \cite{Adhikari1} in the sense that the I-V curves are not strictly pinched in zero, is clearly captured by the model (Fig.~\ref{fig:simulations}). 
Indeed, two mechanisms homogenize the ionic distribution in the absence of electric field: the electrostatic interaction between the ions and the uniform background charge, and the thermal diffusion of the individual ions; in the model, the ratio of these two effects is given by $\tilde T$.
Due to these effects, the memory of the material has a \textit{volatile character}, making RTO material a good candidate for the applications involving volatile memory components. 

Further work has to be carried out in order to elucidate the conditions of creation of non-volatile (or less volatile) memory effects, that are also observed. This involves a large number of parameters such as annealing conditions, pressure and temperature, nature of the electrodes, quality of the surface contact and will be the object of future studies.   For instance, one may consider using porous electrodes with ionic adsorption or intercalation processes. In this case non volatile memory could be stabilized, opening the way to another range of applications.

\section{Conclusion}

In this work we have demonstrated experimentally and confirmed by simulations that the important  ionic electromigration dynamics prevailing in bulk Rb$_2$Ti$_2$O$_5$ at room temperature leads to non-transverse memristive properties.
Because of the finite dimensions on which the ionic migration occurs, two important behaviors exist: one regime where the ions are freely migrating and another one where the ions are accumulating at the edges of the system, which yields a hysteretic current drop, giving rise to the memristive properties. 
Electrostatic interactions and thermal diffusion both contribute to the volatility of the memory effects.
A microscopic model based on ionic migration and diffusion of one-type charged species is able to reproduce the observed non-transverse I-V curves but points to a missing physical phenomenon that still remains to be identified. 



%

\end{document}